# Negative and nonlinear magnetoresistance effect in silicon strip nuclear radiation detector

Fang-cong Wang (王方聪)[1*], Hai-xia Li(李海霞)[2], Hui Guo(郭慧)[1], Xiao-long Fan(范小龙)[1], Zhan-kui Li(李占奎)[2].

School of physical science and technology, Lanzhou University, Lanzhou 730000, China

Institute of Modern Physics, Chinese Academy of Sciences, Lanzhou 730000, China

**Abstract:**

Both negative magnetoresistance and nonlinear magnetoresisitance were observed in silicon strip nuclear radiation detector in room temperature if we applied high magnetic field intensity in different direction. This result is different with former report. We believe this is the coaction of high electric field （Gunn's effect）and high magnetic field , and the consequence is the variation of number of carriers and the carriers' mobility. The weak localization and Landau energy levels also affect the magnetoresistance. Different crystal orientations have different energy band structures, plus Landau energy levels, complex band structures lead complex carriers' mobility. We also found that the magnetoresisitance effect is anisotropy.

## 1. Introduction

Magnetoresistance (MR) effects of non-magnetic materials such as silicon [1-10] have recently attracted a growing research interest due to their physical interests and potential applications in magneto-electronics devices. We reported a finding of extremely large magnetoresistance effect on conventional silicon based PN junction [11] over a wide range of temperatures with temperature control system applied. Its current-voltage (I-V) behaviors under various external magnetic fields obeyed an exponential relationship. Our results indicated that the conventional PN junction was proposed to be used as a multi-functional material based on the interplay between electronics and magnetic response, which was significant for future magneto-electronics in semiconductor industry.

While these days we fabricated silicon strip nuclear radiation detectors [12] which are exactly same as the former devices, and we tested them in room temperature without any temperature control system applied. The device for testing is shown in Fig 1, and all electrodes of it have been connected to the golden fingers.

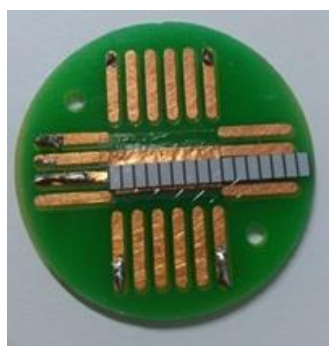

Fig.1. the device (silicon strip nuclear radiation detector) connected to golden fingers for testing (color online).

## 2. Results and Discussion

*Supported by National Natural Science Foundation of China (Grant No.U1531114)

E-mail: wangfc@lzu.edu.cn

The structure of the device is Al/Si (P+)/Si (N)/Si (N+)/Al with a schematic representation shown in Fig. 2. The carrier densities of Si (P+) and Si (N+) were $2.0*10^{14} cm^{-3}$ and $1.0*10^{15} cm^{-3}$, respectively. The device size of P+NN+ junction was designed as 300 μm, much smaller than the several millimeters in previous silicon devices.

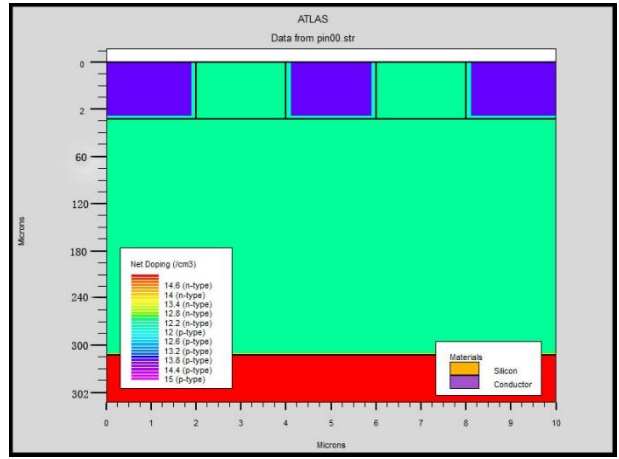

Fig.2. Schematic illustration of the silicon strip nuclear radiation device structure (color online).

We measured silicon strip nuclear radiation detectors resistance and we got the MR-V curves with different B in different direction shown in Fig.3. Two separate Keithly 2400s were used as current meter and voltage source respectively in our experiment.

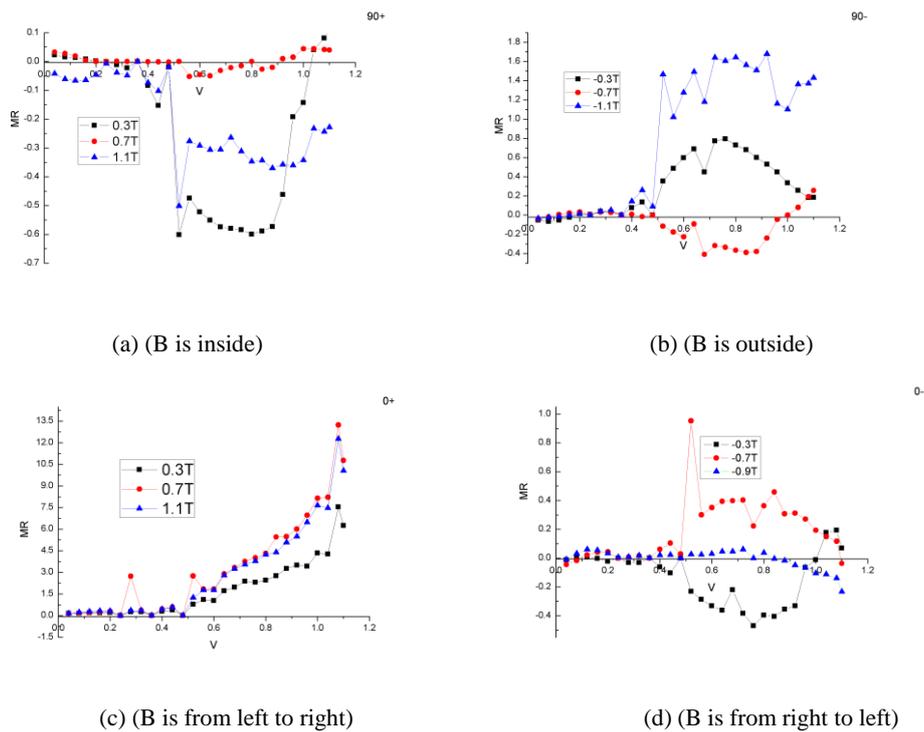

(a) (B is inside)  (b) (B is outside)

(c) (B is from left to right)  (d) (B is from right to left)

Fig.3.The corresponding MR-V curves with different magnetic field intensity in different direction which are all perpendicular to the device (color online).

These results present these four points which are totally different with the former work:

(1) The MR-V curves with various external magnetic field don't obey either linear or exponential relationship;

(2) Negative magnetoresistance is shown in Fig.3a, 3b, 3d;

(3) Saturated magnetoresistance is shown from 0.5V to 1.0V in Fig.3a, 3b, 3d;

(4)Almost all of magnetoresistance curves trend towards 0 when voltage is larger than 1V.

According classical semiconductor physics, the magnetoresistance is proportional to $B^2$, because we test our devices in weak magnetic field at almost all time. For example, we all know that the $\mu_P$=500cm$^2$/V s, so B is weak magnetic field if B is less than 10000G (1T) for P type silicon; also $\mu_N$=1350cm$^2$/V s, B is weak magnetic field if B is less than about 3000G (0.3T) for N type silicon. While in our test environment, the magnetic field changes during 0.3T, 0.7T and 1.1T (or 0.9T) which are uppers or lowers of the boundary of the weak magnetic field and the strong magnetic field. According to the classical theory, when weak B applied to devices, MR is proportional to $B^2$; if continually increasing B, MR is proportional to B; when $B \gg \frac{1}{\mu_H}$ (or $B\mu_H \gg 1$), MR will reach saturation. All of these above classic theory is only considered for one type silicon, while our devices are made of both P type silicon and N type silicon which means different parts of the device will show out different results even if the same magnetic field intensity in same direction is applied to it. So neither linear nor exponential relationship was observed in Fig.3 (a, b, c, d) in silicon strip nuclear radiation detector with the change of magnetic field intensity and the change of direction of it.

We believe the negative magnetoresisitance is the consequence of weak localization [13-15]. Like other hetero structures, quantum coherent scattering of electrons exists in a two-dimensional electron gas system in our PN junction of silicon, and the weak localization effect is the form of the quantum effect in the experiment.

If the a vertical magnetic field is introduced in this two-dimensional electron gas, there will be a phase difference of propagation of the two electron waves along the incoming wave direction and the outgoing wave direction, and the phase difference will destroy the established quantum interference which makes the conductivity back to its previous state, thus the form of weak localization performance is a negative magnetoresistance phenomenon.

In fact, all of these MR effect is only a secondary effect, and the MR is the consequence of the change of the carriers' mobility (Gunn's effect[16]) and the change of number of carriers ($J_S \propto T^{3+\frac{\gamma}{2}}\exp(-\frac{E_g}{k_0 T})$), this means carriers scattering mechanism of semiconductor in different direction is different.

According to the cubic symmetry of the silicon crystals, the minimum value of conduction band is more than one, and there are six valleys of conduction band at several quite symmetric K values in Brillouin zone. We believe the different conduction band in different crystal orientation is the reason of different MR if B applied in different directions. The results of Silicon cyclotron resonance experiment [17] point out that the band structure is complex, N type silicon has one, two or three minimum values (energy valleys) in the conduction band in the different directions: ①when B is applied in [111] direction, there is one absorption peak; ②there are two absorption peaks when B is applied in [110] direction; ③ there are three absorption peaks along any other directions. There are also two absorption peaks observed for P type silicon cyclotron resonance experiments. So different crystal orientations lead different conduction band, then different carriers' mobility in different crystal orientations. The Gunn's effect shown in Fig.4 is that the negative differential mobility occurs if the electrons transfer between two energy valleys of conduction band when E is larger than $E_{th}$. Plus Landau energy levels [18], conduction band becomes much complexer. So Gunn's effect becomes much complexer too.

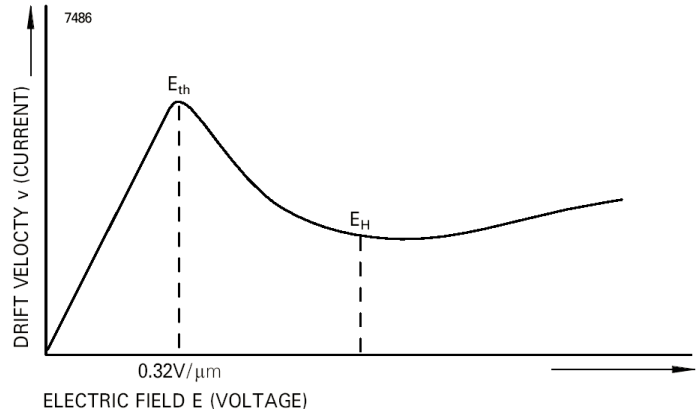

Fig.4. The Gunn's effect occurs in semiconductor if strong electric field applied

When the applied voltage is from 0V to 0.5V which is less than the threshold voltage, no current follows through the device, so $R_B$ almost equals $R_0$, so MR almost equals zero (just like the Fig. 3a, 3b, 3c, 3d). Since there is no current through the device, all of this voltage is applied on the PN junction and there is no voltage on the rest of device, and the electric field on the PN junction is high.

When the voltage is larger than 0.5V, the current begins to follow through the whole device thus the device's temperature increases. The number of carriers will exponential increase with the increase of the temperature. The current should be proportional to the exponential of the voltage. While the electric field on the PN junction must should be between $E_{th}$ and $E_H$, and the $\mu_P$ and $\mu_N$ decrease with the increase of the electric field in Fig.4. The current is $J = (nq\mu_N + pq\mu_P)E$, so the J looks like unchanged with the increase the electric field and saturated magnetoresistance is shown from 0.5V to 1.0V in Fig.3a, 3b, 3d.

While the magnetoresisitance in Fig.3c does not show the saturated magnetoresistance, this exactly shows out that the conduction band is different in different crystal orientation, and there is only one valley and no Gunn's effect in this direction. The MR is anisotropy.

When the voltage is larger than 1V, all the curves of MR in Fig.3 are to be 0. The electric field in device should be greater than the $E_H$ shown in Fig.4, and almost all of the carriers have been pumped into a higher energy valley of the conduction band. The drift velocity (current) becomes minimum, and it does not change with the increase of the electric field, and Ohm's law no longer valid. The electric field and magnetic field have no effect on the current. So the $R_B$ almost equals to the $R_0$, and almost all MR-V curves trend to 0.

## 3. Conclusions:

In summary, negative and nonlinear MR is the consequence of the coaction of high electric field and the high magnetic field: the magnetic field and the electric field both have influence to the magnetoresisitance by changing the number of carriers and changing the carriers' mobility. There are different conduction band structure in different crystal orientation, consequence of applied strong magnetic field is that the band structure plus more Landau energy levels [17] becomes much complexer, so Gunn's effect become much complexer. The carriers' mobility first increase, then decrease, and the velocity remain unchanged with the increase of the electric field. So MR equals to zero when voltage is from 0V to 0.5V. Current follows through the device and the temperature increase, so the number of carriers exponentially increase. While the mobility decreases, so J remains unchanged. So saturated magnetoresistance is just the form of the unchanged J. Keep increasing the voltage, the velocity of carriers remain unchanged, so the almost all MR-V curves trend to zero. When B is no large enough, MR is proportional to $B^2$; Keep increasing B, MR is proportional to B; Keep increasing B, MR remain unchanged. We found that MR showed proportional to voltage only when applied B in some direction. MR could reach to 750% when B was 0.3T and voltage was 1.1V when the B was applied in the same direction.